\documentclass[twocolumn,prb,showpacs]{revtex4}
\usepackage{graphicx}
\usepackage{dcolumn}
\usepackage{bm}
\usepackage{amsmath}

\begin{document}

\preprint{}
\title{ Absence of Hall effect due to Berry curvature in phase space }
\author{Takehito Yokoyama}
\affiliation{Department of Physics, Tokyo Institute of Technology, Tokyo 152-8551, Japan. Email:yokoyama@stat.phys.titech.ac.jp
}
\date{\today}

\begin{abstract}
 Transverse current due to Berry curvature in phase space  is formulated based on the Boltzmann equations with the semiclassical equations of motion for an electron wave packet. 
It is shown that the Hall effect due to the phase space Berry curvature is absent because the contributions from ``anomalous velocity" and  ``effective Lorentz force'' are completely cancelled out.

\end{abstract}

\maketitle


\textit{Introduction.}
Berry phase\cite{Berry} plays an important role in a wide variety of condensed matter physics\cite{Xiao,Vanderbilt} such as electric polarization\cite{Resta,King-Smith}, orbital magnetization\cite{Xiao2,Thonhauser} and magnetoelectric response\cite{Qi,Essin}. The Berry phase is defined as the phase which the eigenstate will pick up when the external parameters of a system form a loop in the parameter space. The paremeters can be momentum, position, and time etc.
Berry curvature in momentum space has a correction to the group velocity of the band dispersion which is perpendicular to the group velocity, anomalous velocity, leading to anomalous Hall effect. \cite{Nagaosa,Liu}
Berry curvature in real space also has a correction to the external electric field (force) which is perpendicular to the electric field, resulting in topological Hall effect. \cite{Ye,Bruno}
Momentum and real space Berry curvatures can be finite when there are nontrivial spin structures in momentum and real spaces, respectively.

The anomalous and topological Hall effects can be realized in several magnets including ferromagnets, antiferromagnets, chiral magnets, and magnetic topological insulators.
Anomalous Hall effect has been observed in, e.g., SrRuO$_3$\cite{Fang}, Mn$_3$Sn,\cite{Nakatsuji} Mn$_3$Ge,\cite{Nayak}  and Cr-doped (Bi, Sb)$_2$Te$_3$\cite{Chang},
while topological Hall effect has been observed in, e.g., Nd$_2$Mo$_2$O$_7$,\cite{Taguchi} MnSi,\cite{Neubauer} MnGe,\cite{Kanazawa} and Mn-doped Bi$_2$Te$_3$ film\cite{Liu2}.
In magnets with strong spin orbit coupling, both anomalous and topological Hall effects can coexist as realized in SrRuO$_3$-SrIrO$_3$ bilayer\cite{Matsuno} and Cr$_x$(Bi$_{1-y}$Sb$_y$)$_{2-x}$Te$_3$/(Bi$_{1-y}$Sb$_y$)$_2$Te$_3$ heterostructures\cite{Yasuda,Jiang}.  
When there are nontrivial spin textures in real and momentum spaces, phase space Berry curvature, namely the Berry curvature including both real and momentum space derivatives, is also finite in general, which may lead to new effects.\cite{Xiao,Xiao2,Xiao3,Zhou} Then, it is natural to ask whether there is a Hall effect stemming from Berry curvature in phase space. This is the problem which we address in this paper.

In this paper, we investigate transverse current due to Berry curvature in phase space  based on the Boltzmann equations with the semiclassical equations of motion for an electron wave packet. 
It is shown that the Hall effect due to the Berry curvature in phase space  is absent because the contributions from ``anomalous velocity'' and  ``effective Lorentz force'' are completely cancelled out.

\textit{Formulation.}
The semiclassical equations of motion for an electron wave packet under an electric field $\bf{E}$ read\cite{Sundaram}

\begin{eqnarray}
\dot {\bf{r}} = {(1 + {\Omega _{{\bf{kr}}}})^{ - 1}}{{\bf{v}}_{\bf{k}}},\\ \hbar \dot {\bf{k}} =  - {(1 - {\Omega _{{\bf{rk}}}})^{ - 1}}e{\bf{E}}
\end{eqnarray}
with the group velocity ${{\bf{v}}_{\bf{k}}} = \frac{1}{\hbar }\frac{{\partial {\varepsilon _{\bf{k}}}}}{{\partial {\bf{k}}}}$, the Berry curvature in phase space  ${\left( {{\Omega _{{\bf{kr}}}}} \right)_{ab}} = i\left( {\left\langle {{{\partial _{{k_a}}}u}}
 \mathrel{\left | {\vphantom {{{\partial _{{k_a}}}u} {{\partial _{{r_b}}}u}}}
 \right. \kern-\nulldelimiterspace}
 {{{\partial _{{r_b}}}u}} \right\rangle  - \left\langle {{{\partial _{{r_b}}}u}}
 \mathrel{\left | {\vphantom {{{\partial _{{r_b}}}u} {{\partial _{{k_a}}}u}}}
 \right. \kern-\nulldelimiterspace}
 {{{\partial _{{k_a}}}u}} \right\rangle } \right)$, and ${\Omega _{{\bf{kr}}}} =  - \Omega _{{\bf{rk}}}^t$. Here, $\left|u \right\rangle$ and $t$ denote the Bloch eigenstate and transpose, respectively.

The distribution function $f$ obeys the Boltzmann equation:\cite{Ashcroft,Sinitsyn}
\begin{eqnarray}
f - {f_0} =  - \tau_{\bf{k}} \hbar \dot {\bf{k}} \cdot {{\bf{v}}_{\bf{k}}}\frac{{\partial {f_0}}}{{\partial \varepsilon }}.
\end{eqnarray}
Here, $f_0$ is the Fermi distribution function, and $\tau_{\bf{k}}$ is the transport lifetime:
\begin{equation}
\frac{1}{\tau_{\bf{k}}}=\sum_{\mathbf{k}^{\prime}} \omega_{\mathbf{k}, \mathbf{k}^{\prime}}\left(1-\cos \theta \right)
\end{equation}
where $\omega_{\mathbf{k}, \mathbf{k}^{\prime}}$  and $\theta$ are the scattering rate and the angle between ${{\bf{v}}_{\bf{k}}}$ and ${{\bf{v}}_{\bf{k}^{\prime}}}$, respectively.

\textit{Results.}
Now, we will show that Hall effect due to phase space Berry curvature is absent. First, consider two dimensional systems. Then, we have from Eqs.(1) and (2)
\begin{eqnarray}
\det (1 + {\Omega _{{\bf{kr}}}})\dot {\bf{r}} = (1 + \mathrm{Tr}{\Omega _{{\bf{kr}}}} - {\Omega _{{\bf{kr}}}}){{\bf{v}}_{\bf{k}}}, \\ \det(1 - {\Omega _{{\bf{rk}}}})\hbar \dot {\bf{k}} =  - (1 - \mathrm{Tr}{\Omega _{{\bf{rk}}}} + {\Omega _{{\bf{rk}}}})e{\bf{E}}
\end{eqnarray}
where ${\Omega _{{\bf{kr}}}}$ and ${\Omega _{{\bf{rk}}}}$ are 2$\times$2 matrices. In the following, for simplicity, we will drop the terms with Tr${\Omega _{{\bf{kr}}}}$ and Tr${\Omega _{{\bf{rk}}}}$, which does not change the conclusions. Note that Tr${\Omega _{{\bf{kr}}}}=-$Tr${\Omega _{{\bf{rk}}}}$ holds. 

The charge current is calculated as 
\begin{widetext}
\begin{eqnarray}
{\bf{j}} =  - e\int {\dot {\bf{r}}fDd{\bf{k}}}  =  - \frac{{e }}{{{{(2\pi )}^2}}}\int {(1 - {\Omega _{{\bf{kr}}}}){{\bf{v}}_{\bf{k}}}(1 + {\Omega _{{\bf{rk}}}})e{\bf{E}} \cdot {{\bf{v}}_{\bf{k}}}\tau_{\bf{k}}\frac{{\partial {f_0}}}{{\partial \varepsilon }}\frac{{d{\bf{k}}}}{{\det (1 + {\Omega _{{\bf{kr}}}})}}}  \nonumber \\
 =  - \frac{{e^2 }}{{{{(2\pi )}^2}}}\int {\left( {({\bf{E}} \cdot {{\bf{v}}_{\bf{k}}}){{\bf{v}}_{\bf{k}}} - ({\bf{E}} \cdot {{\bf{v}}_{\bf{k}}}){\Omega _{{\bf{kr}}}}{{\bf{v}}_{\bf{k}}} + ({{\bf{v}}_{\bf{k}}} \cdot {\Omega _{{\bf{rk}}}}{\bf{E}}){{\bf{v}}_{\bf{k}}} - ({{\bf{v}}_{\bf{k}}} \cdot {\Omega _{{\bf{rk}}}}{\bf{E}}){\Omega _{{\bf{kr}}}}{{\bf{v}}_{\bf{k}}}} \right)\tau_{\bf{k}}\frac{{\partial {f_0}}}{{\partial \varepsilon }}\frac{{d{\bf{k}}}}{{\det (1 + {\Omega _{{\bf{kr}}}})}}} 
\end{eqnarray}
where $D$ is the density of states:
$D = \frac{1}{{{{(2\pi )}^2}}}{\det(1 - {\Omega _{{\bf{rk}}}})}$\cite{Xiao2}.
Thus, we obtain the conductivity tensor of the form
\begin{eqnarray}
{\sigma _{ij}} =  - \frac{e^2}{{{{(2\pi )}^2}}}\int {\left( {{\bf{v}}_{\bf{k}}^i{\bf{v}}_{\bf{k}}^j - \Omega _{{\bf{kr}}}^{ik}{\bf{v}}_{\bf{k}}^k{\bf{v}}_{\bf{k}}^j + \Omega _{{\bf{rk}}}^{kj}{\bf{v}}_{\bf{k}}^k{\bf{v}}_{\bf{k}}^i - \Omega _{{\bf{rk}}}^{kj}\Omega _{{\bf{kr}}}^{il}{\bf{v}}_{\bf{k}}^k{\bf{v}}_{\bf{k}}^l} \right)\tau_{\bf{k}} \frac{{\partial {f_0}}}{{\partial \varepsilon }}\frac{{d{\bf{k}}}}{{\det (1 + {\Omega _{{\bf{kr}}}})}}}
\end{eqnarray}
\end{widetext}
where the superscripts denote the components of vectors and matrices, and the repeated indices are summed over.

We see that the off diagonal conductivity is finite in general and satisfies $\sigma_{ij}=\sigma_{ji}$ since ${\Omega _{{\bf{kr}}}} =  - \Omega _{{\bf{rk}}}^t$. Therefore, the Hall effect is absent. Intuitively, this may be interpreted as follows. 
As for the anomalous Hall effect, a contribution from the Berry curvature to velocity (anomalous velocity) is the origin of this effect. Regarding the topological Hall effect,  a contribution from the Berry curvature to force (effective Lorentz force) leads to this effect.
The phase space Berry curvature induces a velocity perpendicular to the group velocity ${\bf{v}}$ and a force  perpendicular to the electric field ${\bf{E}}$ as seen from Eqs.(5) and (6) (due to the antisymmetric parts of ${\Omega _{{\bf{kr}}}}$ and $\Omega _{{\bf{rk}}}$). Both terms contribute to the Hall response. However, these two contributions completely cancel out each other since ${\Omega _{{\bf{kr}}}} =  - \Omega _{{\bf{rk}}}^t$, resulting in the vanishing of the Hall effect.

Next, consider three dimensional systems. Following the same procedure as in the two dimensional systems, we obtain the conductivity tensor of the form
\begin{eqnarray}
{\sigma _{ij}} =  - \frac{{{e^2}}}{{{{(2\pi )}^3}}}\int {{{(A{{\bf{v}}_{\bf{k}}})}_i}{{(A{{\bf{v}}_{\bf{k}}})}_j}{\tau _{\bf{k}}}\frac{{\partial {f_0}}}{{\partial \varepsilon }}\frac{{d{\bf{k}}}}{{{{\rm det} A} }}}
\end{eqnarray}
with $A \equiv {(1 + {\Omega _{{\bf{kr}}}})^{ - 1}}$.
We again find that the conductivity tensor is symmetric and hence the Hall effect is absent. The ladder vertex corrections are included in the relaxation time in Eq. (4). As can be seen from Eq. (9), even if we multiply any other (vertex) functions by the integrand of  Eq. (9), the Hall effect is still absent.

If fact, Berry curvature in phase space just gives some corrections to the anomalous and topological Hall effects. To see this, let us consider the semiclassical equations of motion for an electron wave packet\cite{Sundaram}
\begin{eqnarray}
\dot{\bf{r}} &={{\bf{v}}_{\bf{k}}} -{\Omega _{{\bf{kr}}}} \dot{\bf{r}}-{\Omega _{{\bf{kk}}}} \dot{\bf{k}}, \\
\hbar \dot {\bf{k}} &=-e {\bf{E}}+{\Omega _{{\bf{rr}}}} \hbar \dot{\bf{r}} +{\Omega _{{\bf{rk}}}}\hbar \dot {\bf{k}}
\end{eqnarray}
where $\Omega_{{\bf{kk}}}$ and $\Omega_{{\bf{rr}}}$ are Berry curvature in momentum and real spaces respectively. These equations reduce to 
\begin{eqnarray}
\dot{\bf{r}}=BA {{\bf{v}}_{\bf{k}}}+\frac{e}{\hbar} \tilde{\Omega}_{\bf{kk}} {\bf{E}},\\
\dot {\bf{k}}= \tilde{\Omega} _{{\bf{rr}}} {{\bf{v}}_{\bf{k}}}-\frac{e}{\hbar} A^t B^t {\bf{E}}
\end{eqnarray}
with $B=\left(1+A {\Omega _{{\bf{kk}}}} A^{t} {\Omega _{{\bf{rr}}}}\right)^{-1}$, $\tilde{\Omega} _{{\bf{kk}}}=BA{\Omega _{{\bf{kk}}}}A^t$ and  $\tilde{\Omega}_{{\bf{rr}}}=A^t{\Omega _{{\bf{rr}}}}B A$. Note that $\tilde{\Omega} _{{\bf{kk}}}$ and  $\tilde{\Omega} _{{\bf{rr}}}$ are antisymmetric: $\tilde{\Omega} _{{\bf{kk}}}^t=-\tilde{\Omega} _{{\bf{kk}}}$ and $\tilde{\Omega} _{{\bf{rr}}}^t=-\tilde{\Omega} _{{\bf{rr}}}$. Therefore, Berry curvature in phase space renormalizes the Berry curvatures in momentum and real spaces, leading to corrections to the anomalous and topological Hall effects. 
When ${\Omega} _{{\bf{rr}}} =0$ or ${\Omega} _{{\bf{kk}}} =0$, we find that Berry curvature effects can be resonantly enhanced when  $\det(1+{\Omega _{{\bf{kr}}}}) \sim 0$.
These results are applicable to ,e.g., chiral magnets which may be described by the Rashba model with inhomogeneous magnetization.

Although Hall effect due to Berry curvature in phase space is absent, it also affects other physical quantities since it gives a correction to the density of states.\cite{Xiao2}

As an example, let us consider the surface state of topological insulators coupled to a magnet where the inplane magnetization varies in one dimension ($x$) as conical magnets. The Hamiltonian is then given by 
\begin{eqnarray}
H = (\hbar v{k_y} + {m_x}(x)){\sigma _x} + ( - \hbar v{k_x} + {m_y}(x)){\sigma _y} + {m_z}{\sigma _z} 
\end{eqnarray}
where ${\sigma _i}$ and $m_i (i=x, y, z)$ represent the Pauli matrices and the exchange field, respectively.
Then, the density of states in the presence of the phase space Berry curvature is calculated as\cite{Xiao} 
\begin{eqnarray}
D  = \frac{1}{{{{(2\pi )}^2}}}\left| {1 - {{\left( {{\Omega _{{\bf{rk}}}}} \right)}_{xx}}} \right| = \frac{1}{{{{(2\pi )}^2}}}\left| {1 - \frac{{\hbar v{m_z}{m_x}'}}{{2\varepsilon _{}^3}}} \right|
\end{eqnarray}
with 
$\varepsilon  =  \pm \sqrt {{{(\hbar v{k_y} + {m_x}(x))}^2} + {{( - \hbar v{k_x} + {m_y}(x))}^2} + m_z^2}$.
The phase space Berry curvature ${{\left( {{\Omega _{{\bf{rk}}}}} \right)}_{xx}}$ has an expression similar to that of the momentum space Berry curvature for massive Dirac fermions\cite{Xiao}. It becomes large when the spatial gradient of the magnetization is large and the energy is around the bottom of the conduction band or top of the valence band. When ${{\left( {{\Omega _{{\bf{rk}}}}} \right)}_{xx}} = 1$, the density of states becomes zero and consequently some physical quantities may become zero or divergent.
For  $v \sim 10^6$m/s, ${m_x}' \sim 1$meV/nm, and $\varepsilon, m_z \sim $10 meV, we have ${{\left( {{\Omega _{{\bf{rk}}}}} \right)}_{xx}} \sim 3$.

\textit{Conclusions.}
In this paper, we have investigated transverse current due to Berry curvature in phase space  based on the Boltzmann equations with the semiclassical equations of motion for an electron wave packet. 
We have shown that the Hall effect due to the Berry curvature in phase space  is absent because the contributions from ``anomalous velocity'' and  ``effective Lorentz force'' are completely cancelled out.

This work was supported by JSPS KAKENHI Grant Number JP30578216 and the JSPS-EPSRC Core-to-Core program ``Oxide Superspin".

\end{document}